%Paper: gr-qc/9506056
%From: ESPOSITO@napoli.infn.it
%Date: Tue, 27 Jun 1995 8:52:20 +0200 (CET-DST)

\magnification \magstep1
\raggedbottom
\openup 4\jot
\voffset6truemm
\def\cstok#1{\leavevmode\thinspace\hbox{\vrule\vtop{\vbox{\hrule\kern1pt
\hbox{\vphantom{\tt/}\thinspace{\tt#1}\thinspace}}
\kern1pt\hrule}\vrule}\thinspace}
\headline={\ifnum\pageno=1\hfill\else
\hfill {\it Spin-raising operators and spin-${3\over 2}$
potentials in quantum cosmology} \hfill \fi}
\rightline {June 1995, DSF report 95/32}
\centerline {\bf SPIN-RAISING OPERATORS AND
SPIN-${3\over 2}$ POTENTIALS}
\centerline {\bf IN QUANTUM COSMOLOGY}
\vskip 0.3cm
\centerline {\bf Giampiero Esposito${ }^{1,2,3}$ and
Giuseppe Pollifrone${ }^{1}$}
\vskip 0.3cm
\centerline {\it ${ }^{1}$Istituto Nazionale di Fisica Nucleare}
\centerline {\it Mostra d'Oltremare Padiglione 20, 80125
Napoli, Italy;}
\centerline {\it ${ }^{2}$Dipartimento di Scienze Fisiche}
\centerline {\it Mostra d'Oltremare Padiglione 19, 80125 Napoli,
Italy;}
\centerline {\it ${ }^{3}$Dipartimento di Fisica Teorica}
\centerline {\it Universit\`a degli Studi di Salerno}
\centerline {\it 84081 Baronissi, Italy.}
\vskip 0.3cm
\noindent
{\bf Abstract.} Local boundary conditions involving field strengths
and the normal to the boundary, originally studied in
anti-de Sitter space-time, have been recently considered in
one-loop quantum cosmology. This paper derives the conditions
under which spin-raising operators preserve these local
boundary conditions on a 3-sphere for fields of spin
$0,{1\over 2},1,{3\over 2}$ and $2$. Moreover, the
two-component spinor analysis of the four potentials of the totally
symmetric and independent
field strengths for spin ${3\over 2}$ is applied to
the case of a 3-sphere boundary. It is shown that such
boundary conditions can only be imposed in a flat
Euclidean background,
for which the gauge freedom in the choice
of the potentials remains.
\vskip 0.1cm
\leftline {PACS numbers: 04.20.Cv, 04.60.+n, 98.80.Dr}
\eject
\noindent
Recent work in the literature has studied the quantization
of gauge theories and supersymmetric field theories in the
presence of boundaries, with application to one-loop
quantum cosmology [1-9]. In particular, in the work described
in [9], two possible sets of local boundary conditions were
studied. One of these, first proposed in anti-de Sitter
space-time [10-11], involves the normal to the boundary
and Dirichlet or Neumann conditions for spin $0$, the normal
and the field for massless spin-${1\over 2}$ fermions, the
normal and totally symmetric field strengths for spins
$1,{3\over 2}$ and $2$. Although more attention has been paid
to alternative local boundary conditions motivated by
supersymmetry, as in [2-3,8-9], the analysis of the former boundary
conditions remains of mathematical and physical interest by
virtue of its links with twistor theory [9]. The aim of
this paper is to derive further mathematical properties of
the corresponding boundary-value problems which are relevant
for quantum cosmology and twistor theory.

In section 5.7 of [9], a flat Euclidean background bounded by
a 3-sphere was studied. On the bounding $S^3$, the following
boundary conditions for a spin-$s$ field were required:
$$
2^{s} \; {_{e}}n^{AA'}...{_{e}}n^{LL'}
\; \phi_{A...L}= \pm {\widetilde \phi}^{A'...L'}
\; \; \; \; .
\eqno (1)
$$
With our notation, ${_{e}}n^{AA'}$ is the Euclidean normal
to $S^3$ [3,9], $\phi_{A...L}=\phi_{(A...L)}$
and ${\widetilde \phi}_{A'...L'}
={\widetilde \phi}_{(A'...L')}$ are totally symmetric
and independent (i.e. not related by any conjugation)
field strengths, which reduce to the massless
spin-${1\over 2}$ field for $s={1\over 2}$. Moreover,
the complex scalar field $\phi$ is such that its real
part obeys Dirichlet conditions on $S^3$ and its imaginary
part obeys Neumann conditions on $S^3$, or the other way
around, according to the value of the parameter
$\epsilon \equiv \pm 1$ occurring in (1), as described
in [9].

In flat Euclidean 4-space, we write the solutions of the
twistor equations [9,12]
$$
D_{A'}^{\; \; \;  (A} \; \omega^{B)}=0
\eqno (2)
$$
$$
D_{A}^{\; \; (A'}\; {\widetilde \omega}^{B')}=0
\eqno (3)
$$
as [9]
$$
\omega^{A}=(\omega^{o})^{A}-i\Bigr({_{e}}x^{AA'}\Bigr)
\pi_{A'}^{o}
\eqno (4)
$$
$$
{\widetilde \omega}^{A'}=({\widetilde \omega}^{o})^{A'}
-i\Bigr({_{e}}x^{AA'}\Bigr){\widetilde \pi}_{A}^{o}
\; \; \; \; .
\eqno (5)
$$
Note that, since unprimed and primed spin-spaces are no longer
isomorphic in the case of Riemannian 4-metrics, Eq. (3) is not
obtained by complex conjugation of Eq. (2). Hence the spinor
field ${\widetilde \omega}^{B'}$ is independent of
$\omega^{B}$. This leads to distinct solutions (4)-(5), where the
spinor fields $\omega_{A}^{o},{\widetilde \omega}_{A'}^{o},
{\widetilde \pi}_{A}^{o},\pi_{A'}^{o}$
are covariantly constant with respect to the
flat connection $D$, whose corresponding spinor covariant
derivative is here denoted by $D_{AB'}$. In section 5.7 of [9]
it was shown that the spin-lowering operator [9,12] preserves
the local boundary conditions (1) on a 3-sphere of radius
$r$ if and only if
$$
\omega_{A}^{o}=-{i \epsilon r \over \sqrt{2}}
\; {\widetilde \pi}_{A}^{o}
\eqno (6)
$$
$$
{\widetilde \omega}_{A'}^{o}=-{i \epsilon r \over
\sqrt{2}} \; \pi_{A'}^{o}
\; \; \; \; .
\eqno (7)
$$
To derive the corresponding preservation condition for
spin-raising operators [12], we begin by studying the
relation between spin-${1\over 2}$ and spin-$1$ fields.
In this case, the independent spin-$1$ field strengths
take the form [9,11-12]
$$
\psi_{AB}=i \; {\widetilde \omega}^{L'}
\Bigr(D_{BL'} \; \chi_{A}\Bigr)
-2\chi_{(A} \; {\widetilde \pi}_{B)}^{o}
\eqno (8)
$$
$$
{\widetilde \psi}_{A'B'}=-i \; \omega^{L}
\Bigr(D_{LB'} \; {\widetilde \chi}_{A'}\Bigr)
-2{\widetilde \chi}_{(A'} \; \pi_{B')}^{o}
\eqno (9)
$$
where the independent spinor fields $\Bigr(\chi_{A},
{\widetilde \chi}_{A'}\Bigr)$ represent a massless
spin-${1\over 2}$ field obeying the Weyl equations
on flat Euclidean 4-space and subject to the boundary
conditions
$$
\sqrt{2} \; {_{e}}n^{AA'} \; \chi_{A}=
\epsilon \; {\widetilde \chi}^{A'}
\eqno (10)
$$
on a 3-sphere of radius $r$. Thus, by requiring that (8)
and (9) should obey (1) on $S^3$ with $s=1$, and bearing
in mind (10), one finds
$$ \eqalignno{
2\epsilon \biggr[\sqrt{2} \; {\widetilde \pi}_{A}^{o} \;
{\widetilde \chi}^{(A'} \; {_{e}}n^{AB')}
-{\widetilde \chi}^{(A'} \; \pi^{o \; B')}\biggr]
&=i \biggr[2 \; {_{e}}n^{AA'} \; {_{e}}n^{BB'} \;
{\widetilde \omega}^{L'} \; D_{L'(B} \; \chi_{A)}\cr
&+\epsilon \; \omega^{L} \; D_{L}^{\; \; (B'} \;
{\widetilde \chi}^{A')}\biggr]
&(11)\cr}
$$
on the bounding $S^3$. It is now clear how to carry out the
calculation for higher spins. Denoting by $s$ the spin
obtained by spin-raising, and defining $n \equiv 2s$,
one finds
$$ \eqalignno{
n \epsilon \biggr[\sqrt{2} \; {\widetilde \pi}_{A}^{o} \;
{_{e}}n^{A(A'} \; {\widetilde \chi}^{B'...K')}
-{\widetilde \chi}^{(A'...D'} \; \pi^{o \; K')}\biggr]
&=i \biggr[2^{n\over 2} \; {_{e}}n^{AA'} ...
{_{e}}n^{KK'} \; {\widetilde \omega}^{L'} \;
D_{L'(K} \; \chi_{A...D)}\cr
&+\epsilon \; \omega^{L} \;
D_{L}^{\; \; (K'} \; {\widetilde \chi}^{A'...D')}
\biggr]
&(12)\cr}
$$
on the 3-sphere boundary. In the comparison spin-$0$ vs
spin-${1\over 2}$, the preservation condition is not
obviously obtained from (12). The desired result is here found
by applying the spin-raising operators [12] to the
independent scalar fields $\phi$ and $\widetilde \phi$
(see below)
and bearing in mind (4)-(5) and the boundary conditions
$$
\phi = \epsilon \; {\widetilde \phi}
\; \; \; \; {\rm on} \; \; \; \; S^{3}
\eqno (13)
$$
$$
{_{e}}n^{AA'}D_{AA'}\phi=-\epsilon \; {_{e}}n^{BB'}D_{BB'}
{\widetilde \phi}
\; \; \; \; {\rm on} \; \; \; \; S^{3}
\; \; \; \; .
\eqno (14)
$$
This leads to the following condition on $S^3$
(cf Eq. (5.7.23) of [9]):
$$ \eqalignno{
0&=i\phi \biggr[{{\widetilde \pi}_{A}^{o}\over \sqrt{2}}
-\pi_{A'}^{o} \; {_{e}}n_{A}^{\; \; A'}\biggr]
-\biggr[{{\widetilde \omega}^{K'}\over \sqrt{2}}
\Bigr(D_{AK'}\phi\Bigr)
-{\omega_{A}\over 2} \; {_{e}}n_{C}^{\; \; K'}
\Bigr(D_{\; \; K'}^{C} \phi \Bigr)\biggr]\cr
&+\epsilon \; {_{e}}n_{(A}^{\; \; \; A'} \;
\omega^{B} \; D_{B)A'} \; {\widetilde \phi}
\; \; \; \; .
&(15)\cr}
$$
Note that, whilst the preservation conditions (6-7) for
spin-lowering operators are purely algebraic, the
preservation conditions (12) and (15) for spin-raising
operators are more complicated, since they also involve
the value at the boundary of four-dimensional covariant derivatives
of spinor fields or scalar fields.
Two independent scalar fields have been
introduced, since the spinor fields obtained by applying
the spin-raising operators to $\phi$ and
${\widetilde \phi}$ respectively are independent as well
in our case.

In the second part of this paper, we focus on the totally
symmetric field strengths $\phi_{ABC}$ and
${\widetilde \phi}_{A'B'C'}$ for spin-${3\over 2}$ fields,
and we express them in terms of their potentials, rather
than using spin-raising (or spin-lowering) operators. The
corresponding theory in Minkowski space-time (and curved
space-time) is described in [13-16], and adapted here to
the case of flat Euclidean 4-space with flat connection $D$.
It turns out that ${\widetilde \phi}_{A'B'C'}$ can then be
obtained from two potentials defined as follows. The first
potential satisfies the properties [13-16]
$$
\gamma_{A'B'}^{C}=\gamma_{(A'B')}^{C}
\eqno (16)
$$
$$
D^{AA'} \; \gamma_{A'B'}^{C}=0
\eqno (17)
$$
$$
{\widetilde \phi}_{A'B'C'}=D_{CC'} \; \gamma_{A'B'}^{C}
\eqno (18)
$$
with the gauge freedom of replacing it by
$$
{\widehat \gamma}_{A'B'}^{C} \equiv \gamma_{A'B'}^{C}
+D_{\; \; B'}^{C} \; {\widetilde \nu}_{A'}
\eqno (19)
$$
where ${\widetilde \nu}_{A'}$ satisfies the positive-helicity Weyl
equation
$$
D^{AA'} \; {\widetilde \nu}_{A'}=0
\; \; \; \; .
\eqno (20)
$$
The second potential is defined by the conditions [13-16]
$$
\rho_{A'}^{BC}=\rho_{A'}^{(BC)}
\eqno (21)
$$
$$
D^{AA'} \; \rho_{A'}^{BC}=0
\eqno (22)
$$
$$
\gamma_{A'B'}^{C}=D_{BB'} \; \rho_{A'}^{BC}
\eqno (23)
$$
with the gauge freedom of being replaced by
$$
{\widehat \rho}_{A'}^{BC} \equiv \rho_{A'}^{BC}
+D_{\; \; A'}^{C} \; \chi^{B}
\eqno (24)
$$
where $\chi^{B}$ satisfies the negative-helicity
Weyl equation
$$
D_{BB'} \; \chi^{B}=0
\; \; \; \; .
\eqno (25)
$$
Moreover, in flat Euclidean 4-space the field strength
$\phi_{ABC}$ is expressed in terms of the potential
$\Gamma_{AB}^{C'}=\Gamma_{(AB)}^{C'}$, independent
of $\gamma_{A'B'}^{C}$, as
$$
\phi_{ABC}=D_{CC'} \; \Gamma_{AB}^{C'}
\eqno (26)
$$
with gauge freedom
$$
{\widehat \Gamma}_{AB}^{C'} \equiv \Gamma_{AB}^{C'}
+D_{\; \; B}^{C'} \; \nu_{A}
\; \; \; \; .
\eqno (27)
$$
Thus, if we insert (18) and (26) into the boundary
conditions (1) with $s={3\over 2}$, and require that
also the gauge-equivalent potentials (19) and (27)
should obey such boundary conditions on $S^3$, we
find that
$$
2^{3\over 2} \; {_{e}}n_{\; \; A'}^{A}
\; {_{e}}n_{\; \; B'}^{B}
\; {_{e}}n_{\; \; C'}^{C}
\; D_{CL'} \; D_{\; \; B}^{L'}
\; \nu_{A}=\epsilon \;
D_{LC'} \; D_{\; \; B'}^{L}
\; {\widetilde \nu}_{A'}
\eqno (28)
$$
on the 3-sphere. Note that, from now on (as already done in
(12) and (15)), covariant derivatives appearing in boundary
conditions are first taken on the background and then
evaluated on $S^3$.
In the case of our flat background, (28) is identically
satisfied since $D_{CL'} \; D_{\; \; \; B}^{L'} \; \nu_{A}$
and $D_{LC'} \; D_{\; \; B'}^{L} \; {\widetilde \nu}_{A'}$
vanish by virtue of spinor Ricci identities [17-18]. In
a curved background, however, denoting by $\nabla$ the
corresponding curved connection, and defining
$\cstok{\ }_{AB} \equiv \nabla_{M'(A}
\nabla_{\; \; \; B)}^{M'} \; , \; \cstok{\ }_{A'B'} \equiv
\nabla_{X(A'} \; \nabla_{\; \; B')}^{X}$,
since the spinor Ricci identities we need are [17]
$$
\cstok{\ }_{AB} \; \nu_{C}=\psi_{ABDC} \; \nu^{D}
-2\Lambda \; \nu_{(A} \; \epsilon_{B)C}
\eqno (29)
$$
$$
\cstok{\ }_{A'B'} \; {\widetilde \nu}_{C'}
={\widetilde \psi}_{A'B'D'C'} \;
{\widetilde \nu}^{D'} -2 {\widetilde \Lambda}
\; {\widetilde \nu}_{(A'} \; \epsilon_{B')C'}
\eqno (30)
$$
one finds that the corresponding boundary conditions
$$
2^{3\over 2} \; {_{e}}n_{\; \; A'}^{A}
\; {_{e}}n_{\; \; B'}^{B}
\; {_{e}}n_{\; \; C'}^{C}
\; \nabla_{CL'} \; \nabla_{\; \; \; B}^{L'}
\; \nu_{A}=\epsilon \; \nabla_{LC'}
\; \nabla_{\; \; B'}^{L}
\; {\widetilde \nu}_{A'}
\eqno (31)
$$
are identically satisfied if and only if one of the
following conditions holds: (i) $\nu_{A}=
{\widetilde \nu}_{A'}=0$; (ii) the Weyl spinors
$\psi_{ABCD},{\widetilde \psi}_{A'B'C'D'}$ and the
scalars $\Lambda,{\widetilde \Lambda}$ vanish everywhere.
However, since in a curved space-time
with vanishing $\Lambda,{\widetilde \Lambda}$, the potentials
with the gauge freedoms (19) and (27) only exist provided
$D$ is replaced by $\nabla$ and the trace-free part
$\Phi_{ab}$ of the Ricci tensor vanishes as well [19],
the background 4-geometry is actually flat
Euclidean 4-space. Note that we require that (31) should
be identically satisfied to avoid that, after a gauge
transformation, one obtains more boundary conditions than
the ones originally imposed. The curvature of the background
should not, itself, be subject to a boundary condition.

The same result can be derived by using
the potential $\rho_{A'}^{BC}$ and its independent
counterpart $\Lambda_{A}^{B'C'}$. This spinor field
yields the $\Gamma_{AB}^{C'}$ potential by means of
$$
\Gamma_{AB}^{C'}=D_{BB'} \; \Lambda_{A}^{B'C'}
\eqno (32)
$$
and has the gauge freedom
$$
{\widehat \Lambda}_{A}^{B'C'} \equiv \Lambda_{A}^{B'C'}
+D_{\; \; A}^{C'} \; {\widetilde \chi}^{B'}
\eqno (33)
$$
where ${\widetilde \chi}^{B'}$ satisfies the positive-helicity
Weyl equation
$$
D_{BF'} \; {\widetilde \chi}^{F'}=0
\; \; \; \; .
\eqno (34)
$$
Thus, if also the gauge-equivalent potentials (24) and (33)
have to satisfy the boundary conditions (1) on $S^3$, one
finds
$$
2^{3\over 2} \; {_{e}}n_{\; \; A'}^{A}
\; {_{e}}n_{\; \; B'}^{B}
\; {_{e}}n_{\; \; C'}^{C}
\; D_{CL'} \; D_{BF'} \;
D_{\; \; A}^{L'} \;
{\widetilde \chi}^{F'}
=\epsilon \; D_{LC'} \; D_{MB'} \;
D_{\; \; A'}^{L} \; \chi^{M}
\eqno (35)
$$
on the 3-sphere. In our flat background, covariant derivatives
commute, hence (35) is identically satisfied by virtue of (25)
and (34). However, in the curved case the boundary conditions
(35) are replaced by
$$
2^{3\over 2} \; {_{e}}n_{\; \; A'}^{A}
\; {_{e}}n_{\; \; B'}^{B}
\; {_{e}}n_{\; \; C'}^{C}
\; \nabla_{CL'} \; \nabla_{BF'}
\; \nabla_{\; \; A}^{L'}
\; {\widetilde \chi}^{F'}
=\epsilon \; \nabla_{LC'} \;
\nabla_{MB'} \; \nabla_{\; \; A'}^{L}
\; \chi^{M}
\eqno (36)
$$
on $S^3$, if the {\it local} expressions of $\phi_{ABC}$ and
${\widetilde \phi}_{A'B'C'}$ in terms of potentials still
hold [13-16]. By virtue of (29)-(30), where $\nu_{C}$ is
replaced by $\chi_{C}$ and ${\widetilde \nu}_{C'}$ is
replaced by ${\widetilde \chi}_{C'}$, this means that
the Weyl spinors $\psi_{ABCD},{\widetilde \psi}_{A'B'C'D'}$
and the scalars $\Lambda,{\widetilde \Lambda}$ should
vanish, since one should find
$$
\nabla^{AA'} \; {\widehat \rho}_{A'}^{BC}=0
\; \; \; \;
\nabla^{AA'} \; {\widehat \Lambda}_{A}^{B'C'}=0
\; \; \; \; .
\eqno (37)
$$
If we assume that
$\nabla_{BF'} \; {\widetilde \chi}^{F'}=0$ and
$\nabla_{MB'} \; \chi^{M}=0$, we have to show that (36)
differs from (35) by terms involving a part of the curvature
that is vanishing everywhere.
This is proved by using the basic rules
of two-spinor calculus and spinor Ricci identities [17-18].
Thus, bearing in mind that [17]
$$
\cstok{\ }^{AB} \; {\widetilde \chi}_{B'}
=\Phi_{\; \; \; \; L'B'}^{AB} \;
{\widetilde \chi}^{L'}
\eqno (38)
$$
$$
\cstok{\ }^{A'B'} \; \chi_{B}
={\widetilde \Phi}_{\; \; \; \; \; \; LB}^{A'B'}
\; \chi^{L}
\eqno (39)
$$
one finds
$$ \eqalignno{
\nabla^{BB'} \; \nabla^{CA'} \; \chi_{B}&=
\nabla^{(BB'} \; \nabla^{C)A'} \; \chi_{B}
+\nabla^{[BB'} \; \nabla^{C]A'} \; \chi_{B} \cr
&=-{1\over 2} \nabla_{B}^{\; \; B'} \;
\nabla^{CA'} \; \chi^{B}
+{1\over 2} {\widetilde \Phi}^{A'B'LC} \; \chi_{L}
\; \; \; \; .
&(40)\cr}
$$
Thus, if ${\widetilde \Phi}^{A'B'LC}$ vanishes, also the left-hand side
of (40) has to vanish since this leads to the equation
$
\nabla^{BB'} \; \nabla^{CA'} \; \chi_{B}
={1\over 2}
\nabla^{BB'} \; \nabla^{CA'} \; \chi_{B}
$. Hence (40) is identically satisfied. Similarly, the
left-hand side of (36) can be made to vanish identically
provided the additional condition $\Phi^{CDF'M'}=0$ holds.
The conditions
$$
\Phi^{CDF'M'}=0
\; \; \; \; \; \; \; \;
{\widetilde \Phi}^{A'B'CL}=0
\eqno (41)
$$
when combined with the conditions
$$
\psi_{ABCD}={\widetilde \psi}_{A'B'C'D'}=0
\; \; \; \;
\Lambda={\widetilde \Lambda}=0
\eqno (42)
$$
arising from (37) for the local existence
of $\rho_{A'}^{BC}$ and $\Lambda_{A}^{B'C'}$ potentials,
imply that the whole Riemann curvature should vanish.
Hence, in the boundary-value problems we are interested in,
the only admissible background 4-geometry (of the Einstein
type [20]) is flat Euclidean 4-space.

In conclusion, in our paper we have completed the characterization
of the conditions under which spin-lowering and spin-raising
operators preserve the local boundary conditions studied in
[9-11]. Note that, for spin $0$, we have introduced a pair of
independent scalar fields on the real Riemannian section of
a complex space-time, following [21], rather than a single scalar
field, as done in [9]. Setting $\phi \equiv \phi_{1}+i\phi_{2},
{\widetilde \phi} \equiv \phi_{3}+i\phi_{4}$, this choice leads
to the boundary conditions
$$
\phi_{1}=\epsilon \; \phi_{3}
\; \; \; \;
\phi_{2}=\epsilon \; \phi_{4}
\; \; \; \; {\rm on}
\; \; \; \; S^{3}
\eqno (43)
$$
$$
{_{e}}n^{AA'} \; D_{AA'} \; \phi_{1}=-\epsilon \;
{_{e}}n^{AA'} \; D_{AA'} \; \phi_{3}
\; \; \; \; {\rm on}
\; \; \; \; S^{3}
\eqno (44)
$$
$$
{_{e}}n^{AA'} \; D_{AA'} \; \phi_{2}=-\epsilon \;
{_{e}}n^{AA'} \; D_{AA'} \; \phi_{4}
\; \; \; \; {\rm on}
\; \; \; \; S^{3}
\eqno (45)
$$
and it deserves further study.

We have then focused on the potentials for
spin-${3\over 2}$ field strengths in
flat or curved Riemannian 4-space bounded
by a 3-sphere. Remarkably, it turns out that
local boundary conditions involving field strengths and
normals can only be imposed in a flat Euclidean background,
for which the gauge freedom in the choice of the
potentials remains. In [16] it was found that $\rho$ potentials
exist {\it locally} only in the self-dual Ricci-flat case,
whereas $\gamma$ potentials may be introduced in the
anti-self-dual case.
Our result may be interpreted as a further restriction provided
by (quantum) cosmology.

A naturally occurring question is whether the potentials studied
in this paper can be used to perform one-loop calculations for
spin-${3\over 2}$ field strengths subject to (1) on $S^3$.
This problem may provide another example (cf [9]) of the fertile
interplay between twistor theory and quantum cosmology, and its
solution might shed new light on one-loop quantum cosmology
and on the quantization program for gauge theories in the presence
of boundaries [1-9].
\vskip 1cm
\leftline {\bf Acknowledgments}
\vskip 1cm
\noindent
The first author is indebted to Roger Penrose for bringing
[13-16] to his attention. The second author is
grateful to the Istituto Astronomico dell'Universit\`a
di Roma for hospitality. Anonymous referees suggested
the correct form of some statements originally formulated
in a misleading way in the first version of our work.
\vskip 1cm
\leftline {\bf References}
\vskip 1cm
\item {[1]}
Moss I G and Poletti S 1990 {\it Nucl. Phys.}
B {\bf 341} 155
\item {[2]}
Poletti S 1990 {\it Phys. Lett.} {\bf 249B} 249
\item {[3]}
D'Eath P D and Esposito G 1991 {\it Phys. Rev.}
D {\bf 43} 3234
\item {[4]}
D'Eath P D and Esposito G 1991 {\it Phys. Rev.}
D {\bf 44} 1713
\item {[5]}
Barvinsky A O, Kamenshchik A Y, Karmazin I P and
Mishakov I V 1992 {\it Class. Quantum Grav.}
{\bf 9} L27
\item {[6]}
Kamenshchik A Y and Mishakov I V 1992
{\it Int. J. Mod. Phys.} A {\bf 7} 3713
\item {[7]}
Barvinsky A O, Kamenshchik A Y and Karmazin I P 1992
{\it Ann. Phys., N.Y.} {\bf 219} 201
\item {[8]}
Kamenshchik A Y and Mishakov I V 1993 {\it Phys. Rev.}
D {\bf 47} 1380
\item {[9]}
Esposito G 1994 {\it Quantum Gravity, Quantum Cosmology
and Lorentzian Geometries} Lecture Notes in Physics,
New Series m: Monographs vol m12
second corrected and enlarged edn
(Berlin: Springer)
\item {[10]}
Breitenlohner P and Freedman D Z 1982 {\it Ann. Phys., N.Y.}
{\bf 144} 249
\item {[11]}
Hawking S W 1983 {\it Phys. Lett.} {\bf 126B} 175
\item {[12]}
Penrose R and Rindler W 1986 {\it Spinors and Space-Time,
Vol. 2: Spinor and Twistor Methods in Space-Time Geometry}
(Cambridge: Cambridge University Press)
\item {[13]}
Penrose R 1990 {\it Twistor Newsletter} {\bf 31} 6
\item {[14]}
Penrose R 1991 {\it Twistor Newsletter} {\bf 32} 1
\item {[15]}
Penrose R 1991 {\it Twistor Newsletter} {\bf 33} 1
\item {[16]}
Penrose R 1991 Twistors as Spin-${3\over 2}$ Charges
{\it Gravitation and Modern Cosmology} eds A Zichichi,
V de Sabbata and N S\'anchez (New York: Plenum Press)
\item {[17]}
Ward R S and Wells R O 1990 {\it Twistor Geometry and
Field Theory} (Cambridge: Cambridge University Press)
\item {[18]}
Esposito G 1993 {\it Nuovo Cimento} B {\bf 108} 123
\item {[19]}
Buchdahl H A 1958 {\it Nuovo Cim.} {\bf 10} 96
\item {[20]}
Besse A L 1987 {\it Einstein Manifolds} (Berlin: Springer)
\item {[21]}
Hawking S W 1979 The path integral approach to quantum gravity
{\it General Relativity, an Einstein Centenary Survey}
eds S W Hawking and W Israel (Cambridge: Cambridge University
Press)
\bye